\documentclass[10pt, conference, letterpaper]{IEEEtran}
\IEEEoverridecommandlockouts
\usepackage{cite}
\usepackage{amsmath,amssymb,amsfonts}
\usepackage{algpseudocode}
\usepackage{algorithmicx}
\usepackage[ruled,vlined,linesnumbered]{algorithm2e}
\usepackage[english]{babel}

\usepackage{graphicx}
\usepackage{textcomp}
\usepackage{xcolor}
\usepackage{subfigure}
\usepackage{amsmath,bm}
\usepackage{amsthm}
\usepackage{float}
\usepackage{stfloats}
\usepackage{url}
\usepackage{multirow}
\usepackage{comment}
\usepackage{CJK}
\usepackage{mathrsfs}
\usepackage{enumitem}
\usepackage{physics}

\usepackage[T1]{fontenc}

\usepackage{array}

\usepackage{cleveref}

\usepackage[letterpaper, left=0.625in, right=0.625in, top=0.75in, bottom=1in]{geometry}

\usepackage{comment}
\def\BibTeX{{\rm B\kern-.05em{\sc i\kern-.025em b}\kern-.08em
    T\kern-.1667em\lower.7ex\hbox{E}\kern-.125emX}}
\begin{document}

\title{\fontsize{20}{0}\selectfont A Performance Analysis Modeling Framework for Extended Reality Applications in Edge-Assisted Wireless Networks\\ \vspace{-0.15 in}
}

\author{
\IEEEauthorblockN{Anik Mallik\IEEEauthorrefmark{1}, Jiang Xie\IEEEauthorrefmark{1}, and Zhu~Han\IEEEauthorrefmark{2}\\
\IEEEauthorblockA{\IEEEauthorrefmark{1}\textit{The Dept. of Electrical and Computer Engineering, The University of North Carolina at Charlotte,} NC 28223, USA}
\IEEEauthorblockA{\IEEEauthorrefmark{2}\textit{The Dept. of Electrical and Computer Engineering, University of Houston,} TX 77004, USA,\\
and \textit{the Dept. of Computer Science and Engineering, Kyung Hee University,} Seoul, South Korea, 446-701\\
Email: amallik@uncc.edu, linda.xie@uncc.edu, zhan2@uh.edu}}

\thanks{\vspace{-0.02 in}This work was supported by funds from Toyota Motor North America and by the US National Science Foundation (NSF) under Grant No. 1910667, 1910891, 2025284, 2107216, 2128368, 2222810, and 2302469, US Department of Transportation, Amazon, and Japan Science and Technology Agency (JST) Adopting Sustainable Partnerships for Innovative Research Ecosystem (ASPIRE) JPMJAP2326.}
}

\IEEEaftertitletext{\vspace{-2.5\baselineskip}}

\maketitle

\begin{abstract}

Extended reality (XR) is at the center of attraction in the research community due to the emergence of augmented, mixed, and virtual reality applications. The performance of such applications needs to be uptight to maintain the requirements of latency, energy consumption, and freshness of data. Therefore, a comprehensive performance analysis model is required to assess the effectiveness of an XR application but is challenging to design due to the dependence of the performance metrics on several difficult-to-model parameters, such as computing resources and hardware utilization of XR and edge devices, which are controlled by both their operating systems and the application itself. Moreover, the heterogeneity in devices and wireless access networks brings additional challenges in modeling. In this paper, we propose a novel modeling framework for performance analysis of XR applications considering edge-assisted wireless networks and validate the model with experimental data collected from testbeds designed specifically for XR applications. In addition, we present the challenges associated with performance analysis modeling and present methods to overcome them in detail. Finally, the performance evaluation shows that the proposed analytical model can analyze XR applications' performance with high accuracy compared to the state-of-the-art analytical models.

\end{abstract}


\vspace{-0.05 in}
\section{Introduction}


In today's era of technological revolution, augmented reality (AR), mixed reality (MR), and virtual reality (VR) are extending their services to every aspect of society and human lives. These applications are considered under one broader concept -- extended reality (XR) \cite{akyildiz2022wireless}. Modern XR devices range from smartphones, tablets, AR glasses, and MR/VR headsets to autonomous and unmanned aerial vehicles. However, they are usually lightweight and equipped with low computational capability to run high-complexity tasks, such as deep learning applications which usually cause a high latency and energy consumption \cite{haoxin2021energy}. To meet the Quality-of-Service (QoS) requirements, cloud and edge computing supported by wireless networks is often adopted to reduce the latency and energy consumption as well as increase the XR service accuracy \cite{theodoropoulos2022cloud}. 


\par So far, the best way to analyze an XR application's performance is through extensive experimental measurement research using testbeds \cite{mallik2022H264, wang2019energy}, which is a time-consuming and laborious process for XR developers and researchers. Moreover, producing datasets of XR applications' performance results is another strenuous task. \textit{A comprehensive performance analysis model can offer the research community a way to avoid all these exhausting data collection and processing tasks.} 

\par However, existing performance analysis methods cannot provide comprehensive insights into an XR service. For instance, end-to-end latency \cite{van2022edge, ko2018wireless} is often considered an important XR performance metric. However, since each segment in the pipeline of running an XR application, such as frame generation, frame conversion, and inference, strongly depends on hardware settings, the complexity of a task, the number of layers in a neural network, and so on, an XR application's end-to-end latency is significantly influenced by these dependencies, but unfortunately, they are rarely considered in existing latency modeling. Therefore, a comprehensive XR performance analysis model should focus on the individual segments of the application pipeline. 


\par The energy consumption of an XR device depends on the end-to-end latency of the application, including the computing latency at an edge server \cite{haoxin2021energy, wang2021you}, and the power consumption during the application run-time. Existing papers propose various ways to model the energy consumption and then minimize the total energy consumption \cite{rajib2017predictive, li2018energy, apicharttrisorn2019frugal}. While some papers break down the total power consumption to hardware levels \cite{leng2019energy}, deeper insights into an XR application's energy consumption can be found in the segments of an XR pipeline. In addition, if heterogeneous wireless networks and devices are involved in an XR application, the energy model should be inclusive of the latency incurred by different networks, devices, and sensors, which is not considered in any existing research.



\par Sensors and devices of different kinds communicate with an XR device in order to transmit control and environmental information, such as the locations of objects, traffic signals, and street map updates \cite{xu2022aoi}. Roadside units, neighboring XR devices and vehicles, and different Internet-of-Things (IoT) devices act as external sensors that establish communication with the candidate XR device \cite{akyildiz2022wireless2}. This communication takes place either directly with the XR device or via an edge server or orchestrator \cite{nasrin2018sharedmec, kuang2019age, wang2019auto, huang2019smart} over a wireless medium. Additionally, these sensors and devices may be located in the same or different sub-networks using different wireless access technologies \cite{lee2008vertical}. The flow of information through these wireless networks and devices can cause delays in the arrival at an XR device, which impacts the freshness of information. How frequently the sensors and devices generate the information and transmit it to the XR device is a way to measure the Age-of-Information (AoI), which is an important performance metric for XR applications. Without maintaining the freshness of information, XR users may get the wrong results at the wrong place and time, which eventually causes a series of effects spreading from nausea to horrible accidents. Furthermore, the mobility of an XR device in heterogeneous networks gives rise to the necessity of both horizontal (within the same sub-network and access technology) and vertical handoffs (different sub-networks and/or access technologies), which are also known as service migration in the domain of edge computing, that can also influence the performance of an XR application. 

\par \textbf{Background:} An XR application's performance can be analyzed based on numerous metrics, such as frame and refresh rates, field-of-view, service accuracy, jitters, system compatibility and integration, 3D alignment, different user experience metrics (e.g., interactivity, ergonomics, heat dissipation, and intuitiveness), network quality, and battery health management. \textit{In this research, the scope of the XR performance analysis consists of latency, energy consumption, and AoI}. Though existing works consider some of the individual segments of the application pipeline presented in this paper, they are not comprehensive enough to explain the performance of an XR service. Moreover, the proposed analysis framework incorporates new determining factors of XR performance, such as diverse computing resources, memory bandwidth, encoding, neural network complexity, and energy conversion. Therefore, the proposed framework is not incremental, but unique. \textit{The focus of this proposed framework is on comprehensiveness in modeling XR performance metrics, which is essential to achieving more accurate performance analysis.}

\par \textbf{Motivations:} In this paper, we argue that a comprehensive performance analysis model is necessary for XR applications, which enables researchers to analyze the performance for both local (i.e., on-device computation) and remote (i.e., cloud or edge-assisted computation) execution of an XR task irrespective of the number or type of sensors or devices and access technologies involved. Consequently, this motivates us toward a performance analysis modeling framework specifically designed for XR applications performing in edge-assisted wireless networks considering latency, energy consumption, and AoI as the performance metrics. To the best of our knowledge, this is the first research work on comprehensive XR performance analysis. The proposed performance analysis modeling framework is exclusive to XR applications due to the specific XR pipeline considered in this paper. This modeling framework can be used by researchers to model the performance of any specific XR applications by extending or modifying the pipeline.


\par \textbf{Challenges:} The latency and energy consumption of an XR device need to be calculated for each individual segment in an XR application pipeline to get a complete insight into the application's performance. The performance of the individual segments depends on the specifications of an XR device, resource allocations, complexity of the task, different parameters set by the applications (e.g., display and encoding), mobility of the device, and wireless channel conditions. Each of these relations is very difficult to present in analytical forms. For example, the computing resource allocation is dictated by both the XR application and the operating system (OS) of a device based on priority and parallel operations. In addition, the computational complexity of a task is determined by its type and scope. Furthermore, the modeling framework should be designed without making unrealistic assumptions that simplify the job but often contradict the field data. Finally, the validation of the proposed analysis model should be done against data collected from performance measurements via a testbed, which is challenging to design to replicate real-world scenarios.


\par \textbf{Contributions:} Our main contributions in this paper are summarized as follows:

\begin{itemize}
    \item \textbf{Performance analysis modeling framework for XR:} We propose a novel performance analysis modeling framework for XR applications performing in edge-assisted wireless networks. First, we introduce the proposed modeling framework along with the considered XR application pipeline in Section \ref{systemModel}. Then we present the proposed modeling of the key performance metrics, namely, latency, energy consumption, and AoI in Sections \ref{LatencyModel}, \ref{EnergyModel}, and \ref{AoIModel}, respectively, along with associated challenges and mitigation techniques. \textit{The uniqueness of the proposed analysis modeling framework lies in its scalability and adaptability to any XR application 
    scenario.}
    \item \textbf{Experimental research on XR performance analysis:} We conduct experimental research on an XR testbed to create a dataset that is used to validate the proposed model and strengthen the model where regression is needed to avoid unrealistic assumptions. The methodology and experimental setup are discussed in Section \ref{sec:experiment}. 
    \item \textbf{Proposed framework's performance evaluation and comparison:} Finally, we evaluate the proposed performance analysis model, validate it using both experimental data and simulation results, and compare it with two other state-of-the-art analysis models (FACT \cite{liu2018edge} and LEAF \cite{wang2022leaf+}) (Section \ref{Performance}). The comparison shows that our proposed model is able to analyze the performance of an XR application with high accuracy.
\end{itemize}

\vspace{-0.05in}
\section{Related Work}
\vspace{-0.05in}
\textbf{Performance analysis of XR:} Performance analysis of XR applications is not a widely researched topic. Existing research works propose performance optimization through analysis models but do not consider the general XR pipeline that can be applied to any XR applications \cite{liubogoshchev2021adaptive, caricato2014augmented, chen2022enhancing}. Benchmark suites and testbeds are also developed for such performance analysis of XR \cite{huzaifa2021illixr}. The performance of AR and VR is analyzed with models for different metrics, such as video bitrate, responsiveness, image noise and rendering, object locations and marker positions, and volumetric data processing performance \cite{sendari2020performance, reed1995virtual}. However, the scope of these analysis models does not include latency, energy consumption, and AoI, which are essential in analyzing the performance of XR services, and is the key focus of this paper.

\par \textbf{Latency analysis:} Latency models for AR/VR applications are presented in several studies \cite{elbamby2019wireless}. Existing works on latency modeling can be divided into two parts: communication and computation latency. Most of these works focus on communication latency in edge-enabled multimedia services \cite{ko2018wireless}, which do not consider propagation delay or path loss. Computation latency is considered in several other research works where the computation capability of a device is modeled as cycles \cite{van2022edge, hoa2023dynamic, mao2017joint, braud2020multipath}. However, our research shows that the available computation resources are a tuple of processing speed (e.g., central processing unit (CPU) frequency), memory size, and allocated resources determined by the application itself and the OS of the device. Lastly, assumptions are made to model an XR application's latency, such as the use of one single server at a time \cite{liu2018edge} without considering service migration. Our research addresses all of these gaps in a new latency analysis model.


\par \textbf{Energy analysis:} The power consumption models proposed for wireless sensor networks, IoT devices, and video streaming applications \cite{wang2006realistic, jano2023modeling, tang2020alleviating} are not applicable to XR devices due to the differences in the application pipeline. These models do not resemble the segments in the pipeline of XR applications, such as frame generation, frame conversion, and frame extraction. Moreover, although research on server placement optimization \cite{trinh2022deep,li2018energy, apicharttrisorn2019frugal, leng2019energy} considers energy modeling, these works do not address XR devices' comprehensive end-to-end energy consumption. However, fortunately, the measurement studies on energy consumption, although arduous, are proven effective in understanding the energy behavior of XR devices \cite{chen2015smartphone, aggarwal2021802, wang2022leaf+}. Our research leverages the benefits of measurement studies toward proposing a new analytical model for energy consumption by an XR application, which will benefit the scientific community to research more on XR services and devices.
\par \textbf{Age-of-Information analysis:} AoI is a relatively new concept that becomes a performance metric in an XR application when control and environmental information is needed to be sent to XR devices \cite{cao2023toward, mallik2024unleashing}. AoI models are presented in \cite{kuang2019age, xu2022aoi, chen2022aoi, chiariotti2021peak}, where assumptions are made on packet processing policies and wireless networks, but they cannot be applied to XR applications that receive the information either directly from sensors or from an edge server. Lastly, AoI is used to optimize the overall performance in several research works \cite{chen2020minimizing, muhammad2021minimizing, zhang2023vehicle}, which also cannot be applied to general XR scenarios due to the lack of scalability. This paper presents a novel AoI model which addresses all the aforementioned research issues.

\section{Overview of the Proposed XR Performance Analysis Modeling Framework} \label{systemModel}
\vspace{-0.05in}

XR applications are diverse in operation and scope. Ranging from virtual reality games to futuristic infotainment systems in autonomous vehicles, XR applications can have different pipelines and features of operation. Therefore, it is difficult to devise a general performance analysis model for XR. To tackle this challenge, the operation and pipeline of the application need to be understood in detail first. This knowledge of the pipeline can further lead to the analysis modeling. Our proposed XR performance analysis framework provides a guideline for enthusiastic researchers in this domain on how to model different performance metrics of an XR application. The performance metrics studied in this paper are end-to-end latency, energy consumption, and AoI. Our research shows that the following steps are essentially helpful in designing such analysis models.

\begin{itemize}
    \item First, the XR application pipeline, including the individual segments, needs to be identified.
    \item Second, the operation of the pipeline needs to be studied. For example, the segments operating sequentially and in parallel need to be distinguished. Moreover, the operations in a data buffer need to be considered in such modeling.
    \item Third, the factors influencing the latency and energy of each segment need to be analyzed. For example, the latency for frame generation depends on the frame rate, frame resolution, data size, and the allocated computing resource and memory bandwidth of the device. Other segments' details are also discussed in later sections.
    \item Fourth, when explicit analytical modeling form is not possible for some segments, other numerical methods, such as regression analysis, are required to find the analytical form. For example, the encoding and decoding latency depends on so many factors (e.g., different configuration parameters) that a direct analytic form is very hard to find.
    \item Fifth, the analytical modeling for AoI of an XR application needs in-depth knowledge of sensors and devices involved in the pipeline and their communication methods. The packet arrival and service rates, propagation models, and path loss models need to be considered if applicable.
    \item Finally, the models' validation needs to be done with appropriate data, preferably collected through experiments. This validation enables researchers to find flaws in the models and fix them accordingly.
\end{itemize}

\par XR applications have complicated pipelines since they contain the attributes of AR, MR, and VR applications. In this research, we have identified the basic components of an XR application that can explain any other XR application with slight modifications. For example, a multiplayer XR game can include database sharing \cite{huh2019xr}, and a vehicular application can consider map data sharing with the cloud and other vehicles -- which are based on the example XR pipeline explained below.
\par
\begin{figure}[t!]
\centerline{\includegraphics[scale=0.35]{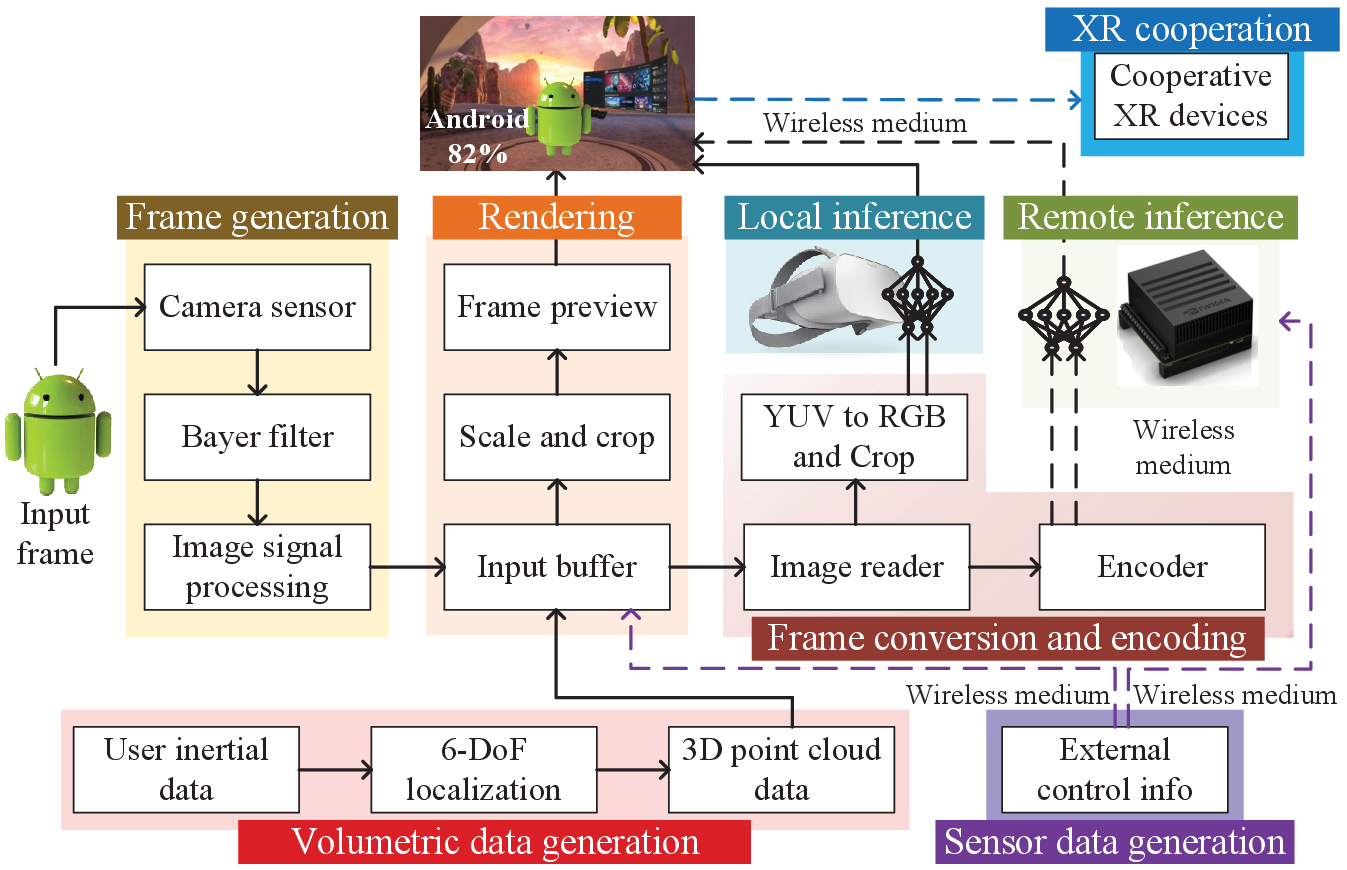}}
\vspace{-0.1 in}
\caption{XR application pipeline for object detection.}
\label{fig:XRpipeline}
\vspace{-0.2 in}
\end{figure}

 To illustrate our proposed modeling framework, in this section, we break down the entire pipeline of an XR application, object detection, into segments that have unique functions and responsibilities. The segments having parallel operations are considered accordingly in the analysis modeling framework. Fig. \ref{fig:XRpipeline} shows these individual segments and an overview of their functions.

\vspace{0.05 in}

\begin{itemize}
    \item \textbf{Frame generation:} The XR device captures a frame from the real world using the camera sensors at a predefined rate (i.e., frame rate). Then, the captured frame is processed by a Bayer filter and signal processor to prepare it for the \textit{input buffer}.
    
    \item \textbf{Volumetric data generation:} The XR device calculates the inertial data of the user (i.e., the person wearing the head-mounted device (HMD) or an autonomous driving system (ADS)). The inertial data is used in the 6 Degrees-of-Freedom (DoF) localization, which is necessary to understand the user's exact location in a three-dimensional (3D) space. Lastly, the 3D point cloud data is extracted from the current scene the XR device is displaying. This data is passed to the \textit{input buffer}.
    
    \item \textbf{External sensor information generation:} In this research, external sensors and devices are considered to be involved in the XR application. The sensors or devices generate and/or provide control information, such as the location of an object or a person in a scene (e.g., a virtual meeting, multiplayer games, and pedestrian location in an ADS). This information is transmitted to the XR device via a wireless medium and stored in the \textit{input buffer} along with the generated frame and volumetric data.
    
    \item \textbf{Frame conversion:} This segment is necessary for the \textit{local inference} (i.e., on-device inference with a light-weight convolutional neural network (CNN)). The raw frames captured in \textit{frame generation} are generally in the color format YUV, which needs to be converted to RGB to match the input format of the CNN. Moreover, the input tensor of the CNN has specific frame resolution requirements which are met through scaling and cropping in this segment. An \textit{image reader} reads the frames from the input buffer and passes them to the \textit{frame converter}.
    
    \item \textbf{Frame encoding:} This segment is related to the \textit{remote inference}. \textit{Frame encoding} compresses the size of the data to be transmitted to an edge server by using predictions and removing redundant features in a frame. In this research, we consider the standard H.264 for the encoding and decoding of the frames. The encoding quality depends on several parameters such as the intervals of I-, P-, and B-frames, bitrate, frame size (i.e., resolution), frame rate (i.e., frames per second (fps)), and quantization value.
    
    \item \textbf{Local inference:} The XR device can detect objects using an on-device light-weight CNN (i.e., CNN with a low number of layers and size), which is used to calculate the inference results without incurring a high delay and energy consumption since the device has a limited computation capability and battery backup. The input from \textit{frame conversion} is directly fed to the \textit{local inference}, and the output of this segment is passed to the \textit{frame renderer}.
    
    \item \textbf{Remote inference:} If the XR device decides to calculate the inference results on an edge server, the \textit{remote inference} is used. In this segment, the encoded frames, along with the volumetric data and control information, are passed to the remote edge server via a wireless medium, which has higher computation resources and a larger CNN to provide more accurate and faster results. The edge server decodes the frame and generates the inference result. The result is then passed to the \textit{frame renderer} via the wireless medium. Note that the computation task can also be distributed between the XR device and edge server(s) (one or multiple servers), based on the application's requirement.
    
    \item \textbf{Frame rendering:} The volumetric data, captured frame, and external control information are used in \textit{frame rendering}. The captured frame is scaled and cropped as per the display requirement of the XR device. In this segment, all of these inputs to the \textit{frame preview} are processed to be displayed accordingly. Finally, the inference results, from either the XR device or the edge server, are passed to the \textit{frame renderer} to display the results alongside.

    \item \textbf{XR cooperation:} The XR device may communicate with other cooperative or collaborative XR devices through edge or cloud servers via a wireless medium. This scenario is common in multi-player gaming or cooperative XR applications. In this segment, the XR device sends either the entire scene or fragments of information to convey specific objects' position and alignment. Generally, this segment is executed alongside frame rendering, hence it is dependent on the application whether to include XR cooperation to the end-to-end latency and energy calculation.
    
\end{itemize}

\section{The End-to-End Latency Analysis Model} \label{LatencyModel}

\subsection{Challenges in Latency Modeling}
An XR application generally has two kinds of latency: computation and communication latency \cite{liu2018edge}. Due to the unique nature and functions of each individual segment of an XR pipeline, analyzing the latency for each of the segments can provide a better insight into the application \cite{wang2022leaf+}. However, the latency of each component depends on several other distinct device and network parameters, which brings additional challenges. For example, most of the computation segments depend on the available computation resources, which cannot be modeled simply. This is because, today's XR devices and edge servers are equipped with modern GPUs. Moreover, the use of tensor processing units (TPUs) and neural processing units (NPUs) is also on the rise. The XR application and the OS of each device together determine which processing units to use and at what utilization ratio.

\par Furthermore, research shows that the depth and size of neural networks (NNs) have impacts on the latency \cite{mallik2023epam}, but the modeling of such impact is not provided in existing works. Exploring the relationship between NNs and the corresponding latency is challenging since each NN is unique in nature in terms of the interconnectivity and use of layers. These challenges are addressed and solved in the following subsection. 

\subsection{The Proposed Latency Analysis Model}

Our proposed latency analysis model consists of the latency from each segment of the XR pipeline for each generated frame. For instance, consider the latency for the $q$-th frame, where $q \in \{1,2,...Q_n\}$ such that $Q_n$ is the final frame at the end of the application. The end-to-end latency of an XR application can be expressed as 
\begin{equation}
\begin{split}
L_{tot}^q\ = & L_{fg}^q + L_{vol}^q + L_{ext}^q + L_{ren}^q + \omega_{loc}L_{fc}^q \\
&+ \bar{\omega}_{loc}L_{en}^q + \omega_{loc}L_{loc}^q + \bar{\omega}_{loc}L_{rem}^q \\
&+ \bar{\omega}_{loc}L_{tr}^q + \bar{\omega}_{loc}L_{HO}^q + L_{coop}^q,
\end{split}
\label{eq:latencyTotal}
\end{equation}\noindent
where $L_{fg}$, $L_{vol}$, $L_{ext}$, $L_{ren}$, $L_{fc}$, $L_{en}$, $L_{loc}$, $L_{rem}$, $L_{tr}$, $L_{HO}$, and $L_{coop}$ are the latency due to frame generation, volumetric data generation, external sensor information generation, frame rendering, frame conversion, frame encoding, local inference, remote inference, transmission, handoff (HO), and XR cooperation. The decision of local inference is denoted as $\omega_{loc}$ and $\bar{\omega}_{loc}$ means the remote inference task, where $\omega_{loc} = \{0,1\}$ is a binary value. Depending on the application, some segments' latency may not need to be incorporated in this model. For example, XR cooperation may be executed in parallel with rendering; therefore, it might be excluded from this calculation.

\par \textbf{Frame generation:} The delay in frame generation depends on the frame rate, frame size, allocated computation resource, data size, and the memory bandwidth of the device, which are denoted as $n_{fps}$, $s_{f1}$, $c_{client}$, $\delta_{f1}$, and $m_{client}$, respectively. The frame size describes the complexity of a task to be processed by the XR device's computation resource. The ratio of $\delta_{f1}$ and $m_{client}$ presents the delay in reading and writing the frame by the memory of the device. Then, the frame generation latency of the $q$-th frame is
\begin{equation}
L_{fg}^q = \frac{1}{n_{fps}^q} + \frac{s_{f1}^q}{c_{client}^q} + \frac{\delta_{f1}^q}{m_{client}^q}.
\label{eq:latencyFG}
\end{equation}\noindent

The values of $n_{fps}$ (frames/s) and $s_{f1}$ (pixel$^2$) are predefined in the application, $\delta_{f1}$ (MB) depends on $s_{f1}$, and $m_{client}$ (GB/s) is a device configuration parameter. 
\par \textit{Computation resource availability:} The XR application requests the OS of a device (e.g., XR device or edge server) to allocate resources of certain processing units at a specific utilization ratio. The OS then allocates the resource, which cannot be expressed in an explicit analytical form easily. Using our collected experimental data, we express the allocated computation resource using multiple linear regression as 
\begin{equation}
\begin{split}
c_{client} = &\omega_c(18.24 + 1.84f_c^2 - 6.02f_c)\\
&+ (1 - \omega_c)(193.67 + 400.96f_g^2 - 558.29f_g),
\end{split}
\label{eq:compClient1}
\end{equation}\noindent
where the processing speeds (i.e., the clock frequency) of CPU and GPU (GHz) are represented as $f_c$ and $f_g$, respectively. The utilization rate of the CPU is denoted as $\omega_{c}$, where $\omega_{c} \in [0,1]$. The GPU utilization rate is $(1 - \omega_{c})$, which means for a task shared by both CPU and GPU, the total resource utilization rate will be equal to $1$. The $R^2$-value of this model is $0.87$, which shows that the regression model is a good fit for the data. Note that this equation can also accommodate the allocation of TPU or NPUs depending on the data availability for proper training of the regression model.

\par \textbf{Volumetric data generation:} The latency due to volumetric data generation depends on the computation resource availability, data size, memory bandwidth, and the virtual scene size. Therefore, the latency for volumetric data generation for the $q$-th frame becomes
\begin{equation}
L_{vol}^q = \frac{s_{vol}^q}{c_{client}^q} + \frac{\delta_{vol}^q}{m_{client}^q},
\label{eq:latencyVol}
\end{equation}\noindent
where $s_{vol}$ and $\delta_{vol}$ are the virtual scene size (pixel$^2$) and the corresponding data size (MB). 

\par \textbf{External sensor information generation:} Assume the external sensors and devices are $m \in \{0,1,...,M\}$, and the XR application requires $n \in \{0,1,...,N\}$, updates during one frame processing time. Denote the $m$-th sensor's latency at the $n$-th update as $L_{ext}^{mn}$. Then the total latency for external sensor information generation for frame $q$ becomes,
\begin{equation}
L_{ext}^q = \max_{m=0}^M{\sum_{n=1}^{N}L_{ext}^{mnq}}.
\label{eq:latencyExt1}
\end{equation}

$L_{ext}^{mnq}$ depends on the information generation frequency of the $m$-th sensor and the propagation delay between the sensor and the XR device. Denote the distance between the $m$-th sensor and the XR device at the $n$-th update during $q$-th frame processing time as $d^{mnq}$ (m). Then $L_{ext}^{mnq}$ becomes
\begin{equation}
L_{ext}^{mnq} = \frac{1}{f_{t}^{m}} + \frac{d^{mnq}}{c},
\label{eq:latencyExt2}
\end{equation}\noindent
where $f_t^m$ and $c$ are the information generation frequency by the $m$-th sensor (Hz) and the propagation speed (m/s), respectively. We assume that there are no path loss, shadowing, or fading effects in this propagation, which can be incorporated into the model according to system requirements.

\par \textbf{Frame rendering:} Frame rendering delay consists of the latency due to computation and file reading and writing. Three types of data are queued in the input buffer: captured frame, volumetric data, and external information, where the associated buffering delays are denoted as $t_f^{buff}$, $t_{vol}^{buff}$, and $t_{ext}^{buff}$, respectively. Then the delay due to data buffering during the $q$-th frame processing time becomes
\begin{equation}
t_{buff}^q = t_f^{buff} + t_{vol}^{buff} + t_{ext}^{buff}.
\label{eq:latencyBuffer}
\end{equation}

Assume data buffering in the input buffer is modeled as a stable $M/M/1$ queueing system. Then the buffering time of these data in the system would be $\frac{1}{\mu - \lambda}$, where $\mu$ is the average service rate and $\lambda$ is the average arrival rate to the buffer with a Poisson distribution. 

Additionally, the frame conversion or encoding and the local or remote inference take place in parallel to the rendering; hence, they are ignored in the calculation of the rendering delay, which contributes to the end-to-end latency, $L_{tot}$. Therefore, the total rendering latency can be expressed as
\begin{equation}
\begin{split}
L_{renTotal}^q = &\frac{s_{f1}^q}{c_{client}^q} + \frac{\delta_{f1}^q}{m_{client}^q} + t_{buff}^q \\
&+ \omega_{loc}L_{tr(loc)}^q + \bar{\omega}_{loc}L_{tr(rem)}^q,
\label{eq:latencyRendering}
\end{split}
\end{equation}\noindent
where $L_{tr(loc)}$ and $L_{tr(rem)}$ are the latency for transmission of local and remote inference results to the renderer, respectively.

\par \textbf{Frame conversion:} The latency of color conversion, scaling, and cropping of frames depends on the computational resources of the device. Therefore, the frame conversion latency during frame $q$'s processing time can be written as
\begin{equation}
L_{fc}^q = \frac{s_{f1}^q}{c_{client}^q} + \frac{\delta_{f1}^q}{m_{client}^q}.
\label{eq:latencyConversion}
\end{equation}

\par \textbf{Frame encoding:} As mentioned earlier, frame encoding delay is dependent on the encoding scheme's different parameters, such as the intervals of I-frame and B-frame, bitrate, frame size, frame rate, and quantization value, which are denoted here as $n_i$, $n_b$, $n_{bitrate}$, $s_{f1}$, $n_{fps}$, and $n_{quant}$, respectively. However, the relation between the encoding latency and these parameters is challenging to express in an explicit analytical form. Instead, we use multiple linear regressions to find the encoding latency at the $q$-th frame as
\begin{small}
\begin{equation}
\begin{split}
L_{en}^q = &\left(-574.36 - 7.71n_i + 142.61n_b + 53.38n_{bitrate} + 1.43s_{f1}^q\right. \\
&+ 163.65n_{fps} + 3.62n_{quant})/c_{client}^q + \frac{\delta_{f1}^q}{m_{client}^q}.
\label{eq:latencyEncoding}
\end{split} \vspace{-0.15in}
\end{equation}    
\end{small}
We include ${\delta_{f1}^q}/{m_{client}^q}$ in this model since the frames need to be read from the input buffer before encoding. The $R^2$-value of this regression model is $0.79$, which shows a good fitness of the model.

\par \textbf{Local inference:} The local inference latency model is a function of the computation complexity of a task, allocated computation resources, data writing and reading delays, and the complexity of the CNN. Hence, the local inference computation latency for frame $q$ can be modeled as \vspace{-0.05in}
\begin{equation}
L_{loc}^q = \omega_{client}\left[\frac{s_{f2}^q}{c_{client}^q \cdot C_{CNN(loc)}} + \frac{\delta_{f2}^q}{m_{client}^q}\right],
\label{eq:latencyLocal}
\vspace{-0.05in}
\end{equation}\noindent
where $s_{f2}$ and $\delta_{f2}$ are the converted frame size and data size, respectively. Due to the activity of the image reader, ${\delta_{f2}^q}/{m_{client}^q}$ is added in this model. The complexity of the lightweight CNN stored on the XR device is denoted by $C_{CNN(loc)}$. Another variable $\omega_{client}$ determines the portion of the split task to the XR device, where $\omega_{client} \in [0,1]$.

\vspace{0.05in}
\par \textit{CNN model complexity in XR performance analysis:} A CNN model's complexity depends on two parameters: the depth (the number of layers) and the size (occupied storage space on the device memory) of the CNN \cite{mallik2023epam, tu2023energy}. Generally, an XR application uses a pre-trained CNN model. Therefore, the complexity of a CNN model is only considered for the inference tasks, not the training. \textit{Note that this CNN complexity is exclusive to XR applications' latency and energy analysis. It is not applicable to other applications.} Furthermore, efficient CNN models are equipped with depth scaling nowadays \cite{wang2023yolov7}, which is also addressed in this research. We use linear regression to find the complexity of a CNN model as
\begin{equation}
C_{CNN} = 2.45 + 0.0025d_{CNN} + 0.03s_{CNN} + 0.0029d_{scale},
\label{eq:complexityCNN}
\end{equation}\noindent
where $d_{CNN}$, $s_{CNN}$, and $d_{scale}$ are the depth, size, and depth scaling factor of a CNN model, respectively. The $R^2$-value of this model is $0.844$. We train the regression model with a vast dataset of different CNN models' latency and energy consumption data which is later discussed in Section \ref{sec:experiment}.

\par \textbf{Remote inference:} Remote inference in an XR application pipeline takes care of two tasks: decoding the received frame and running inference on the decoded frame. Therefore, the latency due to remote inference on a single edge server for the $q$-th frame can be modeled as
\begin{equation}
L_{rem}^q = \omega_{edge}\left[\frac{s_{f3}^q}{c_\epsilon^q \cdot C_{CNN(rem)}} + \frac{\delta_{f3}^q}{m_{\epsilon}^q} + L_{dec}^q\right],
\label{eq:latencyRemote}
\end{equation}\noindent
where $s_{f3}$ and $\delta_{f3}$ are encoded frame size and data size, $c_\epsilon$ and $m_\epsilon$ are allocated computational resources and memory bandwidth of the edge server, $C_{CNN(rem)}$ is the complexity of the large CNN model(s) running on the edge server, and $\omega_{edge}$ represents the portion of the split task to the edge server.

\par Decoding is usually faster than encoding due to the straightforward reconstruction of videos, typically with the help of GPUs and dedicated hardware decoders, while encoding has to deal with many issues, such as removing redundancies and predicting frames. Through our experiments, the decoding delay is found to be around one-third of the encoding delay if conducted on the same device. The percentage of the encoding delay equal to the decoding delay is defined as the discount rate, $\gamma$, in this paper. Therefore, the decoding delay for the same frame can be expressed as
\begin{equation}
L_{dec} = \frac{L_{en} \cdot c_{client} \cdot \gamma}{c_\epsilon}.
\label{eq:decodingDelay}
\end{equation}

Using the experimental data, we derive a relation between the computational resources of the XR device and edge server from this equation as $c_\epsilon = 11.76c_{client}$.

\par \textit{Remote inference on multiple edge servers:} An XR application can split the inference task on multiple edge servers that execute the task in parallel. Let $E$ be the set of edge servers Hence, the latency due to remote inference on multiple edge servers at frame $q$ becomes
\begin{equation}
L_{rem}^q = \max_{e=1}^E\left[\omega_{edge}^eL_{rem}^{eq}\right],
\label{eq:latencyRemoteMultiple}
\end{equation}\noindent
where, $\omega_{edge}^e$ denotes part of the inference task assigned to the $e$-th edge server. Here, $\omega_{client} + \sum_{e=1}^{E}\omega_{edge}^e = \omega_{task}$ where $\omega_{task}$ expresses the total inference task required by the XR application for a single frame. Each edge server's latency, $L_{rem}^{eq}$ at frame $q$ can be modeled according to (\ref{eq:latencyRemote}).

\par \textbf{Transmission latency:} In the XR application pipeline, all data coming from and to the edge server are transmitted via a wireless medium. The transmission latency, therefore, is a function of the data rate and propagation delay, which at frame $q$ can be expressed as 
\vspace{-0.1in}
\begin{equation}
L_{tr}^q = \frac{\delta_{f3}^q}{r_{w}^q} + \frac{d_\epsilon^q}{c},
\label{eq:latencyTransmission}
\vspace{-0.05in}
\end{equation}\noindent
where $r_{w}$ and $d_\epsilon$ are available wireless resources (network throughput in Mbps) and the distance between the edge server and the XR device (m). Wireless path loss is not considered in this model but can be added to the framework based on system requirements. 

\par \textbf{Latency due to handoff (HO):} 
In the proposed framework, we consider an XR device leaving a wireless coverage zone toward another with the same or different access technology, with the mobility of the XR device modeled by the Random walk model. The probability that an XR device moves from one wireless coverage zone to another, $P(HO)$, can be derived using methods in existing papers such as \cite{kim2000dynamic}. In addition, we consider a vertical HO for XR applications and find the HO latency, $l_{HO}$ using a similar analysis as in \cite{xie2007ieee, solouk2011vertical}. The average total HO latency during frame $q$'s processing time is \vspace{-0.05in}
\begin{equation}
L_{HO}^q = l_{HO} \cdot P(HO).
\label{eq:latencyHandoff}
\end{equation}\noindent

\par \textbf{XR cooperation latency:} XR cooperation takes place between the client XR device and other cooperative devices via a wireless medium. Generally, this segment is executed in parallel with rendering, which is why the latency due to XR cooperation does not need to be considered in the end-to-end latency of an XR application. However, this parallel operation is entirely dependent upon the application's scope. Application developers or researchers can still use the following model to separately evaluate an XR application's performance during cooperation or include this latency into the overall latency calculation if needed. The latency due to XR cooperation is expressed as
\begin{equation}
L_{coop}^q = \frac{\delta_{f4}^q}{r_{w}^q} + \frac{d_{coop}^q}{c},
\label{eq:latencyCoop}
\end{equation}\noindent
where $\delta_{f4}$ and $d_{coop}$ represent the data size to be transmitted to the cooperative XR device and the distance between the two communicating devices.


\section{The Energy Consumption Analysis Model} \label{EnergyModel}

\subsection{Challenges in Energy Modeling}
The energy consumption analysis model for XR applications is not as straightforward as measuring the power consumption only. Before proposing an energy model, there has to be sufficient insight into the power consumption behavior of an XR device during the application. For example, the power consumption trend is not the same for all the XR devices running the same application. To tackle this challenge, an analysis of a huge experimental power consumption dataset is required.

\par Additionally, the energy from the battery of an XR device is not entirely used for the XR application. Electrical energy is usually converted to thermal energy by a small percentage. Moreover, no matter whether the XR application is running or not, there is always a small amount of power consumption on the XR device, known as base energy. Therefore, proposing an accurate energy model for XR applications is a challenging task.

\subsection{The Proposed Energy Consumption Analysis Model}

Our proposed energy consumption analysis model for an XR device while running an XR application follows a similar procedure to the latency model described in Section \ref{LatencyModel}. The total energy consumption by an XR device for the $q$-th frame can be expressed as
\begin{equation}
\begin{split}
E_{tot}^q\ =& \left.E_{fg}^q + E_{vol}^q + E_{ext}^q + E_{ren}^q + \omega_{loc}E_{fc}^q\right. \\
&+ \bar{\omega}_{loc}E_{en}^q + \omega_{loc}E_{loc}^q + \bar{\omega}_{loc}E_{rem}^q \\
&+ \bar{\omega}_{loc}E_{tr}^q + \bar{\omega}_{loc}E_{HO}^q + E_{coop}^q + E_\theta^q + E_{base}^q,
\end{split}
\label{eq:energyTotal}
\end{equation}
\noindent
where $E_{fg}$, $E_{vol}$, $E_{ext}$, $E_{ren}$, $E_{fc}$, $E_{en}$, $E_{loc}$, $E_{rem}$, $E_{tr}$, $E_{HO}$, $E_{coop}$, $E_\theta$, and $E_{base}$ represents energy consumption by the XR device during frame generation, volumetric data generation, external information generation, frame rendering, frame conversion, frame encoding, local inference, remote inference, transmission, HO, cooperation, energy converted to thermal energy, and base energy, respectively. We can further express (\ref{eq:energyTotal}) as

\begin{small}
\begin{equation}
\begin{split}
E_{tot}^q\ =& \left(\int_{0}^{L_{fg}^q}P_{fg}^q\,dt + \int_{0}^{L_{vol}^q} P_{vol}^q\,dt + \int_{0}^{L_{ext}^q} P_{ext}^q\,dt \right. \\
&+ \int_{0}^{L_{ren}} P_{ren}^q\,dt + \omega_{loc}\int_{0}^{L_{fc}^q}P_{fc}^q\,dt + \bar{\omega}_{loc}\int_{0}^{L_{en}^q}P_{en}^q\,dt \\
&+ \omega_{loc}\int_{0}^{L_{loc}^q} P_{loc}^q\,dt + \bar{\omega}_{loc}\int_{0}^{L_{rem}^q}P_{rem}^q\,dt \\
&+ \bar{\omega}_{loc}\int_{0}^{L_{tr}^q}P_{tr}^q\,dt + \int_{0}^{L_{HO}^q}P_{HO}^q\,dt + \int_{0}^{L_{coop}^q}P_{coop}^q\,dt \\ 
&+ E_\theta^q + E_{base}^q,
\end{split}
\label{eq:energyPower}
\vspace{-0.08in}
\end{equation}
\end{small}\noindent
where $P_{fg}$, $P_{vol}$, $P_{ext}$, $P_{ren}$, $P_{fc}$, $P_{en}$, $P_{loc}$, $P_{rem}$, $P_{tr}$, $P_{HO}$, and $P_{coop}$ represents power consumed by the XR device during frame generation, volumetric data generation, external information generation, frame rendering, frame conversion, frame encoding, local inference, remote inference, transmission, HO, and XR cooperation. We find that power consumption is a function of the computational resources allocated for the task. The relationships between power and computational resources, base power, and heat dissipation in an XR device are discussed below.

\par \textbf{Computing resource dependent power consumption:} Power consumption during an XR application in an XR device is a function of the computational resources allocated for the task. However, there has been no mathematical formula to establish this relationship. As a result, we find the mean power consumption model of an XR application based on multiple linear regressions using our collected experimental dataset as
\begin{equation}
\begin{split}
P_{mean} =& \left.\omega_c(18.85f_c - 3.64f_c^2 -20.74) +\right. \\
&(1 - \omega_c)(187.48f_g - 135.11 f_g^2 -62.197).
\label{eq:powerComputation}
\end{split}
\end{equation}\noindent

$P_{mean}$ is a function of both CPU and GPU resource utilization. The $R^2$-value of this model in (\ref{eq:powerComputation}) is $0.863$. Interested researchers can extend this relation to accommodate TPU and NPU resources as well.

\par \textbf{Base power in an XR device:} The base power is defined as the small portion of the total power consumption that is always consumed during an XR application due to the minimal background activities and leakage current in the device. The minimal background activities are run by the OS itself, such as the system clock, display, and connectivity functions. Moreover, the leakage current is a semiconductor property that flows even if the XR device is idle.  These background activities and the leakage current gives rise to a small amount of energy consumption throughout an XR application, which is denoted as $E_{base}$ in this paper.

\par \textbf{Heat dissipation:} Heat dissipation in an XR device causes serious discomfort among the users. This heat is generated by the CPU, GPU, and battery of an XR device. A small portion of the total energy consumption is converted into thermal energy and consequently dissipates heat, which is represented by $E_{\theta}$ in (\ref{eq:energyTotal}) and (\ref{eq:energyPower}).

 

\section{The Age-of-Information (AoI) Analysis Model} \label{AoIModel}

\subsection{Challenges in AoI Modeling} \vspace{-0.05in}
In the XR application scenario presented in Fig. \ref{fig:XRpipeline}, the XR device is connected to a number of external sensors and devices that provide the XR device with control and environmental information, as well as scene and other data for cooperative XR. These sensors and devices are \textit{heterogeneous} in characteristics, i.e., different information generation frequencies, different arrival rates to the input buffer, and different service rates by the buffer for each of the information packets. This makes mathematical modeling of AoI difficult.

\subsection{The Proposed AoI Analysis Model} 
\begin{figure}[hbt!]
\centerline{\includegraphics[scale=0.35]{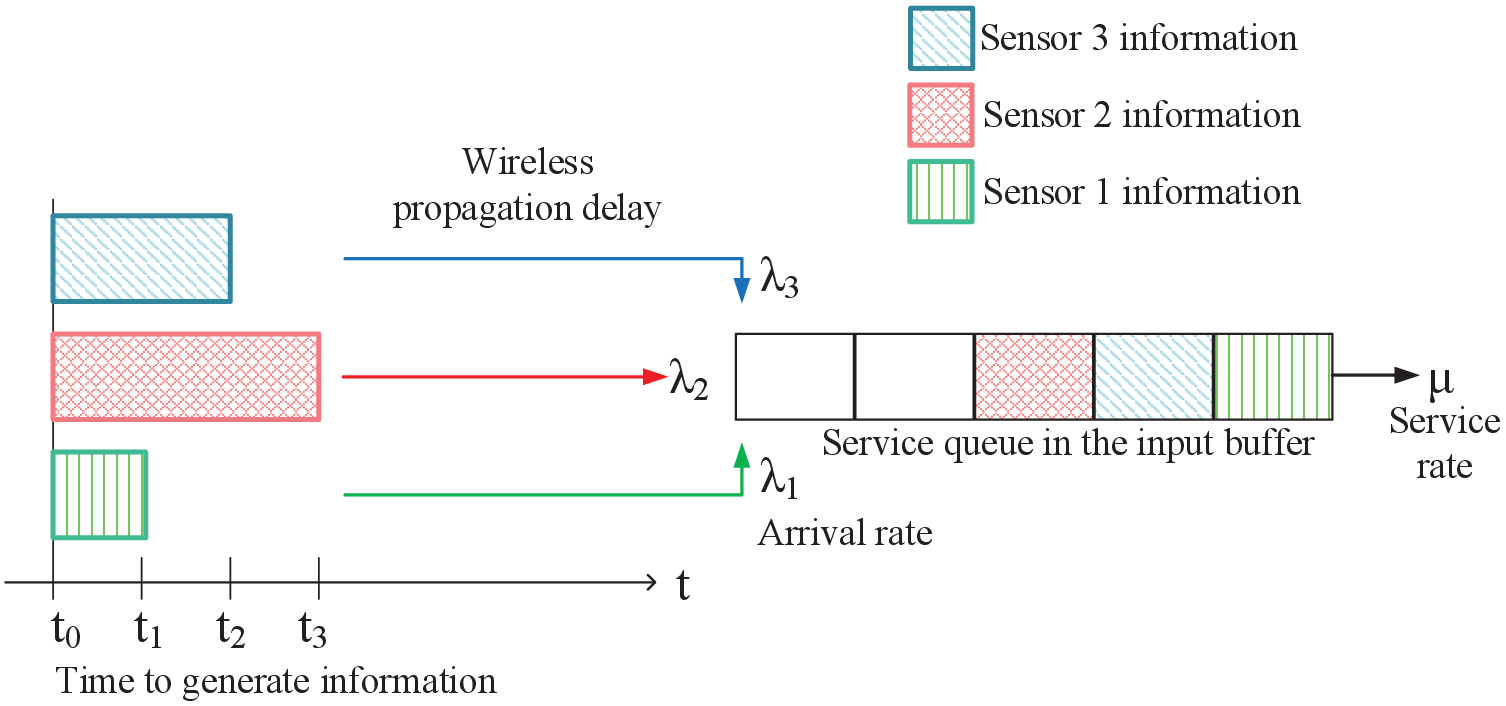}}
\vspace{-0.1 in}
\caption{External sensor information generation, transmission, and service process for XR.}
\label{fig:AoI-XR}
\end{figure}

\vspace{-0.05in}
In this research, a sample scenario is provided in Fig. \ref{fig:AoI-XR}, where three sensors are generating and transmitting information. At time $t=0$, the sensors start generating the information. However, the information generation by all the sensors is not finished at the same time, $t=1$. This can happen due to different information generation frequencies of different sensors and devices. This frequency of the $m$-th sensor or device is denoted by $f_t^m$, where $m\in M$ and $M$ is the set of external sensors and devices connected to the XR device. Note that the packet length is the same for all the sensors in this scenario.

\vspace{-0.05in}

\par \textbf{Average service time and AoI:} The information packets arrive at the input buffer with different arrival rates, $\lambda$, due to differences in $f_t$. If the service rate at the buffer is $\mu$, considering an $M/M/1$ stable system, the average time spent in the buffer by each information packet is
\begin{equation}
\bar{T} = \frac{1}{\mu-\lambda}.
\label{eq:stableT}
\vspace{-0.1in}
\end{equation}

\par The information packets experience propagation delay through the wireless medium, as well. Therefore, the AoI (in seconds) for the $m$-th sensor at the $q$-th frame is
\begin{equation}
t^{mnq} = T^{mnq} + \left(\frac{d_m^q}{c} + \bar{T}\right) - T_{Req}^{nq},
\label{eq:AOIm}
\end{equation}\noindent
where $d_m$ is the distance between the $m$-th sensor and the XR device, $c$ is the propagation speed, $T^{mn}$ is the generation time of the information originated by the sensor for the $n$-th cycle of information generation, and $T_{Req}^n$ is the time requested by the XR device to generate information for the $n$-th cycle. We do not consider the wireless resources (i.e., bandwidth and TCP throughput) in this case since the control information packets are very small. Now, the average AoI of the $m$-th sensor at frame $q$'s processing time is \vspace{-0.1in}
\begin{equation}
A^{mq} = \frac{1}{N}\sum_{n=1}^Nt^{mnq},
\label{eq:averageAOI}
\end{equation}

\par \textbf{Relevance-of-Information (RoI):} We introduce a new metric for XR applications in this section, which is Relevance-of-Information (RoI). RoI (no unit) is defined as the ratio of information generation frequency by an external sensor or device to the information generation frequency required by an XR application to avoid having out-of-date information in a frame. If RoI$\geq 1$, then the information can be considered as fresh. The frequency of information processed by the XR device from the $m$-th sensor is
\begin{equation}
\bar{f}_t^m = \frac{1}{A^{mq}}.
\label{eq:freqAOIm}
\end{equation}

The required information generation frequency is $f_{req}^m = N/L_{tot}^q$, where $N$ is the total number of information updates during the total processing time of frame $q$. Then RoI can be expressed as
\begin{equation}
RoI = \frac{\bar{f}_t^m}{f_{req}^m}.
\label{eq:ROI}
\end{equation}


\section{Experimental Setup and Methodology} \label{sec:experiment}

\begin{figure}[htb!]
\centerline{\includegraphics[scale=0.55]{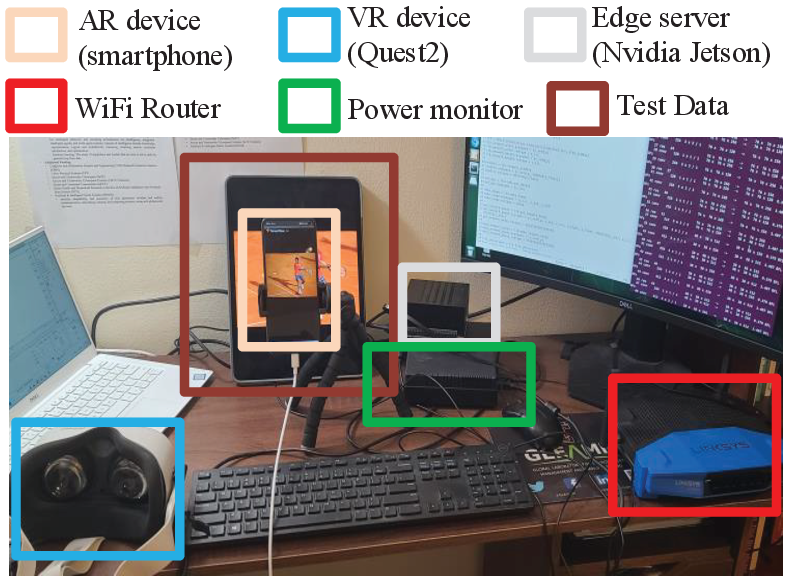}}
\vspace{-0.1 in}
\caption{A snippet of the experimental testbed used in this research.}
\label{fig:Testbed}
\vspace{-0.2 in}
\end{figure}

We implement an XR application on the devices with diverse hardware configurations listed in Table \ref{table:XRdeviceSpecs}. For energy measurement, an external tool named ``Monsoon Power Monitor'' is used that provides power to XR devices with a data sampling rate of once at every $0.2$ ms. Due to the delicate input power terminal design in these devices, the power measurement becomes challenging. Applying a heat gun and soldering process with careful investigation of the power terminals mitigates this problem. All the experiments are carried out in similar environmental conditions to make the experiments controllable and repeatable. A simple configuration of the testbed used in this research is illustrated in Fig. \ref{fig:Testbed}.

\begin{table*}[t!]
\setlength{\tabcolsep}{2.5pt}
\caption{Brief specifications of the XR and edge devices used in the experiments}
\vspace{-0.15in}
\begin{center}

\begin{tabular}{c||cccccccc}
\hline
\textbf{Denotation}   & \textbf{Model} & \textbf{System-on-} & \textbf{CPU}          & \textbf{GPU}          & \textbf{RAM} & \textbf{OS}          & \textbf{Wi-Fi}       & \textbf{Release} \\
\textbf{}             & \textbf{}      & \textbf{Chip}    & \textbf{}             & \textbf{}        & \textbf{}    & \textbf{}            & \textbf{}     & \textbf{Date}    \\ \hline  

XR 1 & Huawei & Kirin 9000 & 8-core (1$\times$3.13GHz A77 & Mali G78 & 8GB & Android 10 & 802.11 & October,\\
& Mate 40 Pro & (5 nm) & 3$\times$2.54GHz A77 & & LPDDR5 & & a/b/g/n/ac/ax & 2020\\
& & & 4$\times$2.05GHz A55) & & & & &\\

XR 2 & OnePlus & Snapdragon & 8-core (1$\times$2.84GHz & Adreno 650 & 8GB & Android 10 & 802.11 & April,\\
& 8 Pro & 865 (7 nm) & 3$\times$2.42GHz & & LPDDR5 & & a/b/g/n/ac/ax & 2020\\
& & & 4$\times$1.8GHz Kryo 585) & & & & & \\

XR 3 & Motorola & Helio P70 & 8-core (4$\times$2.0GHz A73 & Mali G72 & 4GB & Android 9 & 802.11 & October,\\
& One Macro & (12 nm) & 4$\times$2.0GHz A53) & & LPDDR4X & & b/g/n & 2019\\

XR 4 & Xiaomi & Snapdragon & 8-core (4$\times$2GHz Gold & Adreno 610 & 4GB & Android 10 & 802.11 & August,\\
& Redmi Note8 & 665 (11nm) & 4$\times$1.8GHz Silver) Kryo260 & & LPDDR4X & & a/b/g/n/ac & 2020\\

XR 5 & Google Glass & Snapdragon & 8-Core Kryo & Adreno 615 & 3GB & Android & 802.11 & May,\\ 
& Enterprise & XR1 & (2$\times$2.52GHz,& & LPDDR4 & 8.1 & a/g/b/n/ac & 2019\\
&  Edition 2& & 6$\times$1.7GHz) & & & & \\

XR 6 & Meta & Snapdragon & 8-core  & Adreno 650 & 6GB & Oculus OS & 802.11 & October,\\ 
& Quest 2 & XR2 & Kryo 585 & & LPDDR5 & & a/g/b/n/ac/ax& 2020\\

External & Nvidia & Tegra & 2-Core NVIDIA Denver2 & 256-core & 8GB & Ubuntu 18.04 & -- & March,\\ 
\& XR 7& Jetson TX2 & & 4-Core A57 MPCore & NVIDIA Pascal & LPDDR4 & & & 2017\\

Edge & Nvidia & Tegra & 8-core & 512-core & 32GB & Ubuntu 18.04 & -- & October,\\ 
server& Jetson AGX & & ARM v8.2 & Volta GPU with & LPDDR4X & LTS aarch64 & & 2018\\
& Xavier & & &  Tensor Cores\\
\hline
\end{tabular}
\label{table:XRdeviceSpecs}
\end{center}
\vspace{-0.25in}
\end{table*}

\par We use $11$ CNN models with distinct architectures in this research that are listed in Table \ref{table:cnn}. A LinkSys dual-band router (2.4 GHz and 5 GHz) is used as the wireless medium, which connects the XR devices, external sensors, and the server via Wi-Fi. For the remote inference, we use {\fontfamily{qcr}\selectfont YOLOv3} and {\fontfamily{qcr}\selectfont YOLOv7} in the edge server. The 2017 COCO test dataset is used for testing the XR application for both local and remote inferences. 

\begin{table}[b!]
\caption{CNNs used in this research}
\vspace{-0.05in}
\begin{tabular}{l||c|c|c}
\hline
\multicolumn{1}{c||}{\multirow{2}{*}{\textbf{CNN}}} & \multirow{2}{*}{\textbf{\begin{tabular}[c]{@{}c@{}}Model depth\\ (no. of layers)\end{tabular}}} & \multirow{2}{*}{\textbf{\begin{tabular}[c]{@{}c@{}}Storage\\ space (MB)\end{tabular}}} & \multirow{2}{*}{\textbf{\begin{tabular}[c]{@{}c@{}}GPU\\ support\end{tabular}}} \\
\multicolumn{1}{c||}{} &   &  &  \\
\hline
MobileNetv1\_240 Float & 31 & 16.9 & Yes \\
MobileNetv1\_240 Quant & 31 & 4.3 & \\
MobileNetv2\_300 Float & 99 & 24.2 & Yes \\
MobileNetv2\_300 Quant & 112 & 6.9 & \\
MobileNetv2\_640 Float & 155 & 12.3 & Yes \\
MobileNetv2\_640 Quant & 167 & 4.5 & \\
EfficientNet Float & 62 & 18.6 & Yes \\
EfficientNet Quant & 65 & 5.4 & \\
NasNet Float & 663 & 21.4 & Yes \\
YoLoV3 & 106 & 210 & Yes \\
YoLoV7 & Scaling (1.5) & 142.8 & Yes\\                             \hline
\end{tabular}
\label{table:cnn}
\end{table}

\par \textbf{Regression model training:} After conducting experiments, we collect a huge dataset for the design and evaluation of our proposed XR performance analysis modeling framework. To design and evaluate the framework, we use datasets containing $119,465$ and $36,083$ data, respectively. The regression models are trained with the data collected from devices XR1, XR3, XR5, and XR6, and tested with the data collected from XR2, XR4, and XR7. Training and testing with separate datasets help evaluate the models' performance. All the regression-based models used in this research are generated using a $95\%$ confidence boundary.

\vspace{-0.05in}
\section{Performance Evaluation} \label{Performance}
\vspace{-0.05in}
We evaluate our proposed XR performance analysis model's performance compared to the test dataset collected during experiments, which are considered ``Ground Truth (GT)'' in this paper. First, we present the performance evaluation of our proposed model for end-to-end latency and energy consumption of the XR pipeline. Then we evaluate the AoI model's performance based on an emulated experiment. Finally, we compare the end-to-end latency and energy consumption analysis in different conditions by the proposed model with several state-of-the-art analysis models. We evaluate the proposed models' performances for each individual segment of the XR application's pipeline considered in this research. However, due to space constraints, we discuss some of the major evaluation results in this section.
\vspace{-0.05in}
\subsection{End-to-End Latency Validation} \label{LatPerf}
\vspace{-0.05in}


We compare the end-to-end latency of the XR application calculated by our proposed analytical model with the ground truth for both local (Fig. \ref{fig:latloc}) and remote inference (Fig. \ref{fig:latrem}). In remote inference, device mobility is not considered.


\par \textbf{Insights:} A mean error of $2.74\%$ and $3.23\%$ in local and remote inference latency calculation is observed as compared to the ground truth, which means the proposed model is very accurate in calculating the end-to-end latency for XR applications. We find that the more diverse the training dataset for regression is, the more accurate the model can be.

\vspace{-0.05 in}
\subsection{End-to-End Energy Consumption Validation} \label{EnerPerf}
\vspace{-0.05 in}
The comparison of end-to-end energy consumption of the XR service obtained from the proposed model and ground truth is shown in Figs. \ref{fig:enerLoc} and \ref{fig:enerRem} for local and remote inference, respectively.


\par \textbf{Insights:} The mean errors for local and remote inference energy calculation are $3.52\%$ and $5.38\%$, respectively, as compared to the ground truth. This error percentage can be reduced by improving the regression model's performance of the power consumption model.

\vspace{-0.05 in}
\subsection{Age-of-Information Validation} \label{AoiPerf}
\begin{figure*}[t]
\centering
\subfigure[]
{\includegraphics[width=0.3\textwidth]{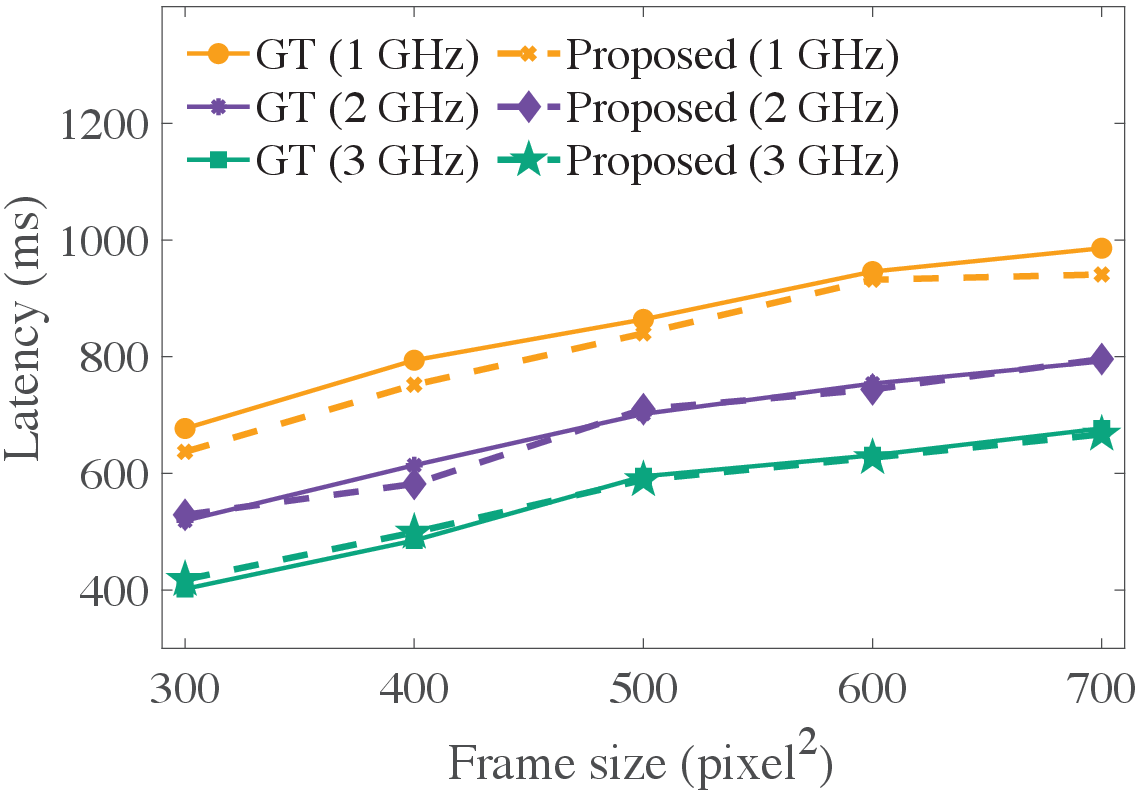}\vspace{-0.05in}
\label{fig:latloc}}
\subfigure[]
{\includegraphics[width=0.3\textwidth]{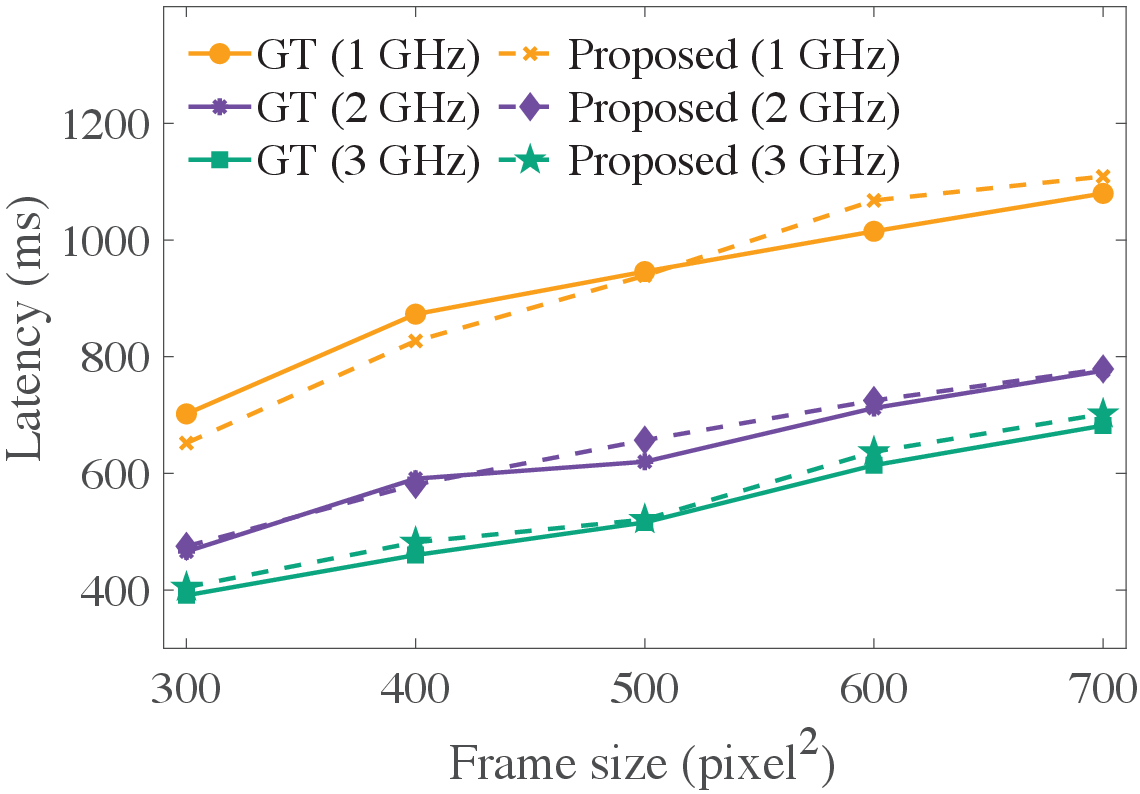}\vspace{-0.05in}
\label{fig:latrem}}
\subfigure[]
{\includegraphics[width=0.3\textwidth]{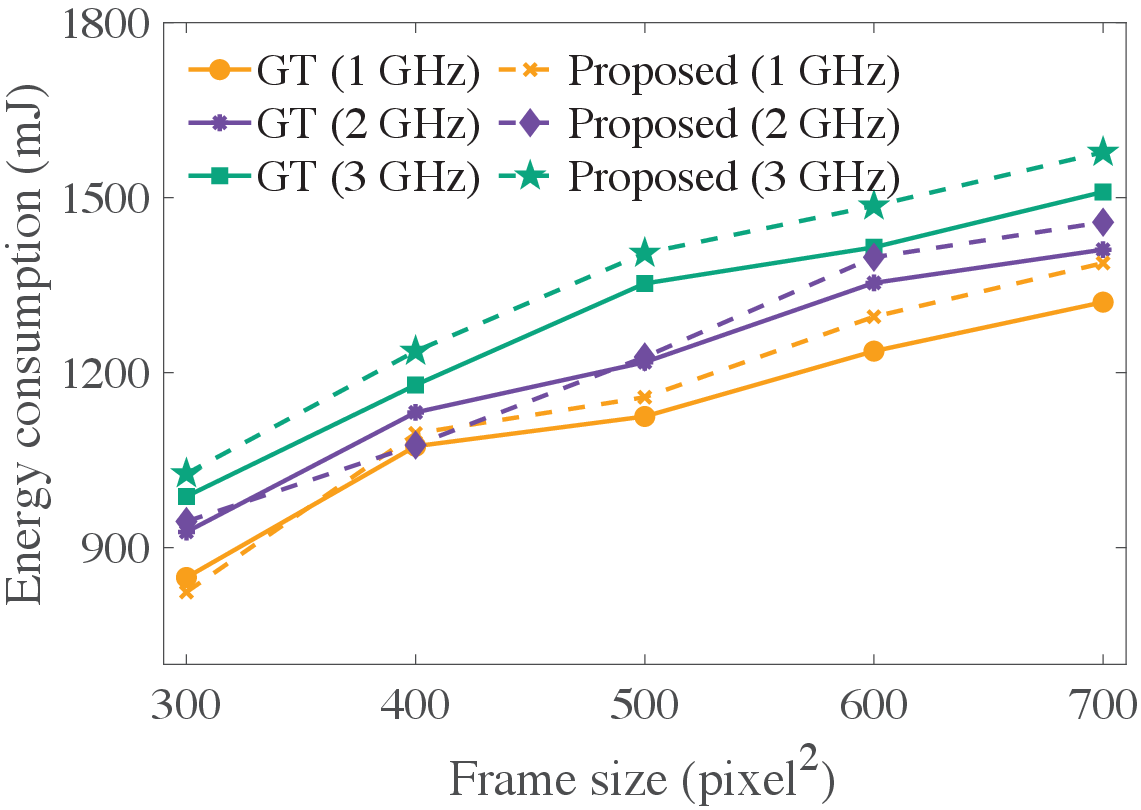}\vspace{-0.05in}
\label{fig:enerLoc}}
\vspace{-0.05in}
\subfigure[]
{\includegraphics[width=0.3\textwidth]{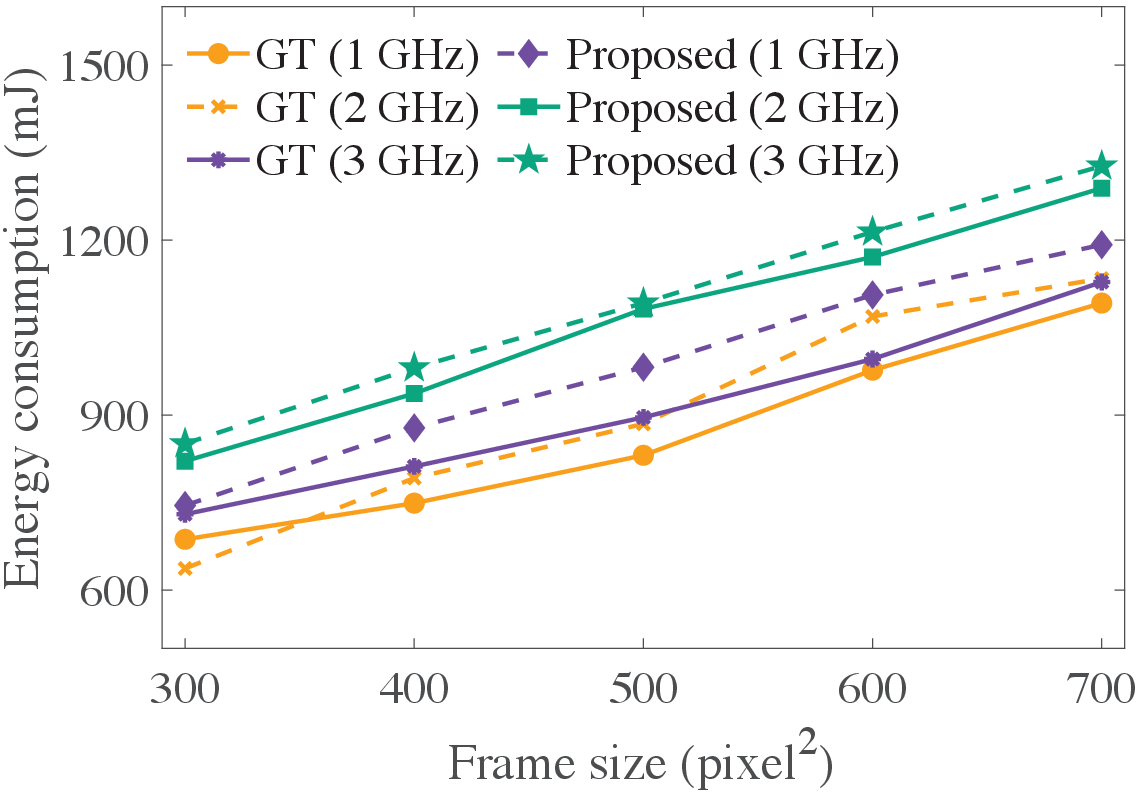}\vspace{-0.05in}
\label{fig:enerRem}}
\subfigure[]
{\includegraphics[width=0.3\textwidth]{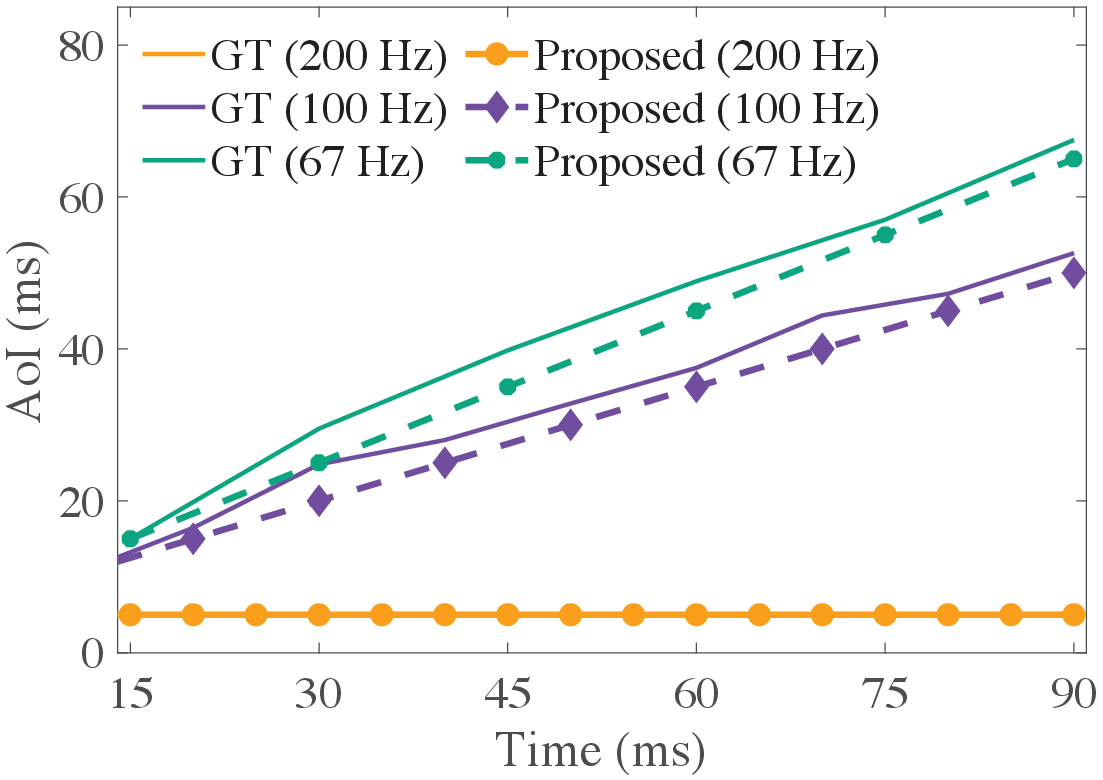}\vspace{-0.05in}
\label{fig:AoIperf}}
\subfigure[]
{\includegraphics[width=0.29\textwidth]{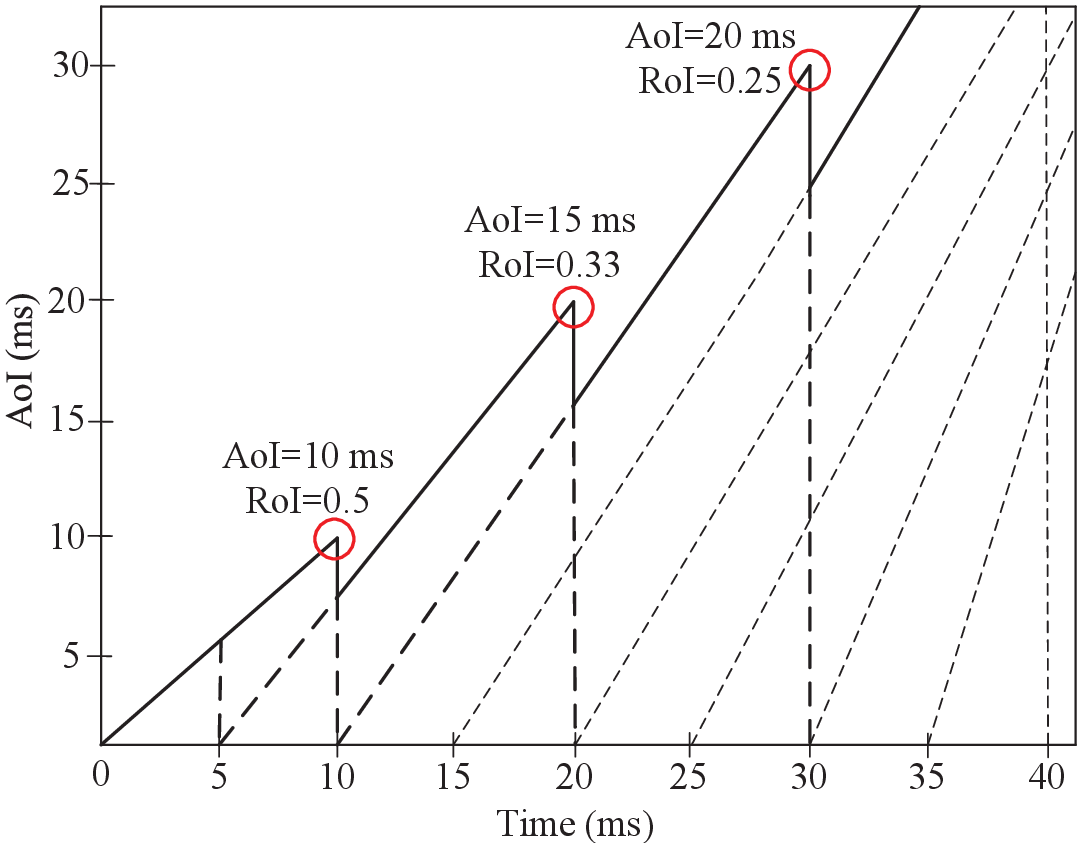}\vspace{-0.05in}
\label{fig:roi}}
\vspace{-0.1in}
\caption{Evaluation of the proposed XR performance analysis models: analysis of end-to-end latency for (a) local and (b) remote execution, end-to-end energy consumption for (c) local and (d) remote execution, AoI at different (e) information generation frequency and (f) RoI for info. gen. frequency of $100$ Hz.}
\label{fig:evaluation}   
\vspace{-0.2in}
\end{figure*}

\vspace{-0.05 in}
A sample AoI scenario is emulated where three sensors have a rate of information generation of 1 every 5, 10, and 15 ms (200 Hz, 100 Hz, and 66.67 Hz), respectively. The XR application has a requirement of 1 information update every 5 ms. From Fig. \ref{fig:AoIperf}, an increase in AoI is observed when the rate of generating information gets lower (i.e., the larger value of frequency). The reason behind this increase is the increase in the delay between information request and information generation at every update cycle (e.g., the sensor generating information at the 67 Hz frequency is transmitting the first information when the third update is required). This incident is explained in detail in Fig. \ref{fig:roi}, where the sensor with a $100$ Hz information generation frequency has incremental AoI at every XR information update period. The corresponding RoI is also shown in the figure.

\par \textbf{Insights:} To maintain a proper AoI, sensors should follow the RoI and use a necessary frequency of information generation.

\subsection{Comparison of Model Performance} \label{Comparison}
\vspace{-0.05 in}

We compare the latency and energy of our proposed model with two existing analytical models: FACT \cite{liu2018edge} and LEAF \cite{wang2022leaf+}. \textit{The reason behind considering these two models is that these are the most comprehensive and accurate state-of-the-art analytical models presented by researchers so far for augmented reality applications -- a subset of XR applications. No other existing models provide better insights into such applications. Therefore, these two models are the best candidates for comparison with our proposed performance analysis framework.}
\begin{itemize}
    \item \textbf{FACT \cite{liu2018edge}:} FACT proposes to include computation, core network, and wireless latency into the overall service latency model of an edge-assisted AR application. The respective energy model can be found by following each component of the service latency. However, FACT presents the computation latency as a function of the computation complexity and available computation resources, which are formulated without considering different processing sources, data size, and the memory of the device, but they are taken into account in this research.
    \item \textbf{LEAF \cite{wang2022leaf+}:} LEAF overcomes several limitations of FACT by breaking down the entire pipeline of an edge-AR application and considering each segment's latency separately. However, it still suffers from the simplicity in formulating the computation latency and energy as FACT does. Our proposed framework presents a comprehensive way to model XR latency and energy consumption to achieve more accurate performance results. 
\end{itemize}

\begin{figure}[t!]
\centering
\subfigure[]
{\includegraphics[width=0.235\textwidth]{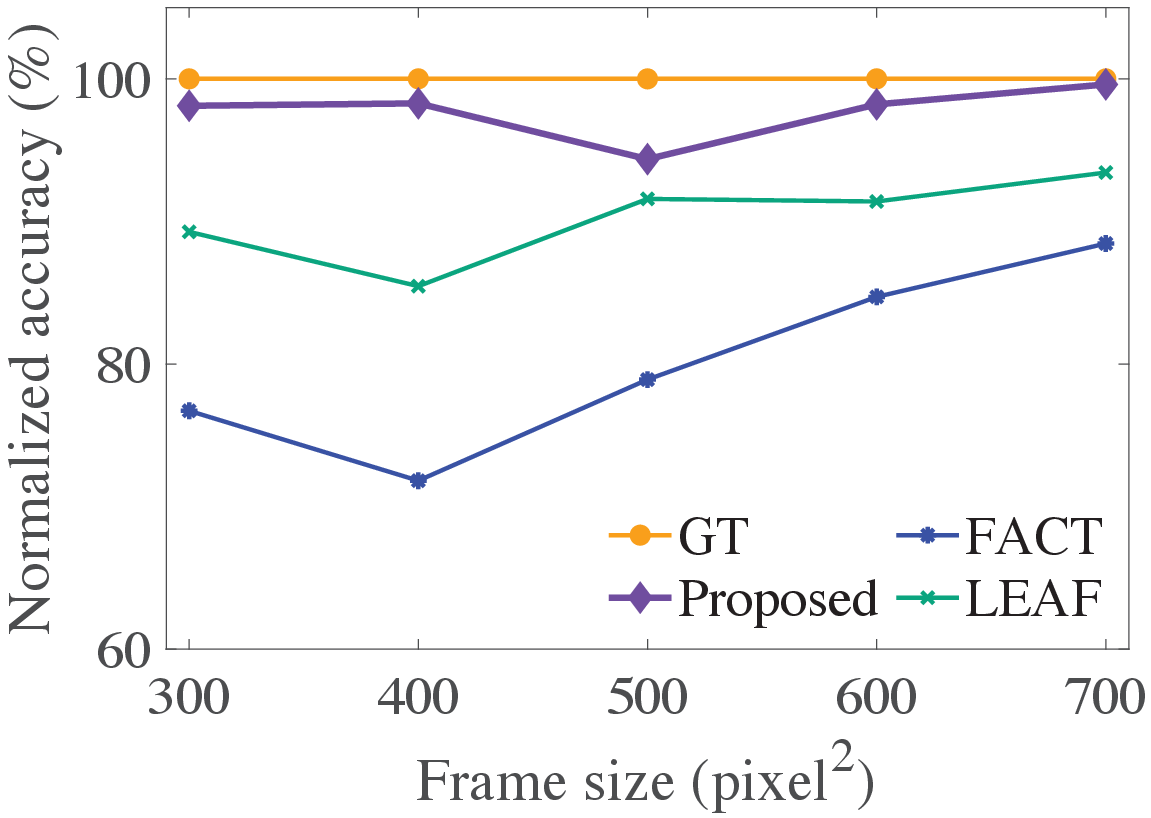}\label{fig:compLat}}
\hspace*
{\fill}
\subfigure[]
{\includegraphics[width=0.24\textwidth]{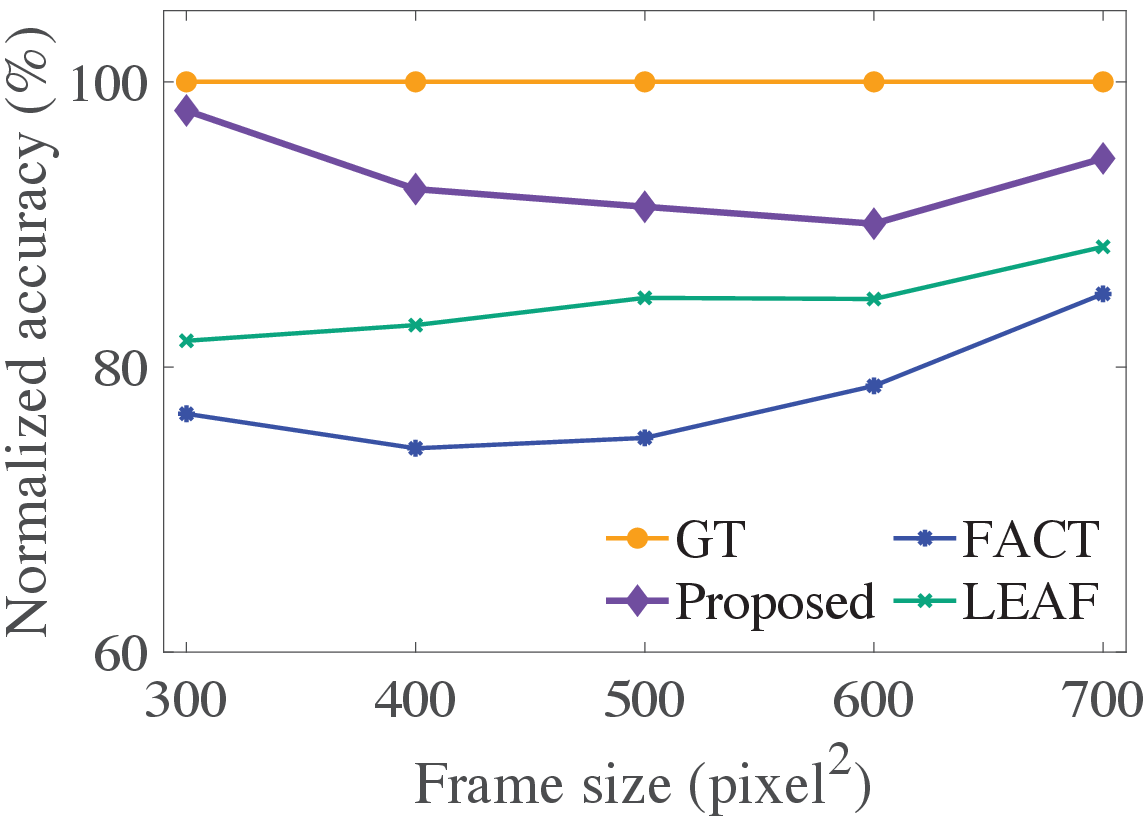}\label{fig:compEner}}
\vspace{-0.15in}

\caption{Comparison of (a) end-to-end latency and (b) end-to-end energy consumption for remote inference obtained from GT and analytical models with FACT \cite{liu2018edge} and LEAF \cite{wang2022leaf+} in terms of normalized accuracy.}
\label{fig:LatEnerAll}   
\vspace{-0.15in}
\end{figure}

The performance comparisons for end-to-end latency and energy consumption using remote inference are shown in Figs. \ref{fig:compLat} and \ref{fig:compEner} in terms of normalized accuracy, where the normalized accuracy of GT is $100\%$. Our proposed analytical model performs with higher accuracy than FACT and LEAF in latency calculation by $17.59\%$ and $7.49\%$, and in energy calculation by $15.30\%$ and $8.71\%$, respectively.

\par \textbf{Insights:} Our proposed XR performance modeling framework achieves this superior performance due to the consideration of the complex models of computation resource, encoding, and transmission, and the relation between the computation resource of the XR device and edge server.

\vspace{-0.05 in}
\section{Conclusion}
\vspace{-0.05 in}
In this paper, we presented a novel modeling framework for performance analysis of XR applications in edge-assisted wireless networks. Our proposed framework consists of analytical methods to evaluate the performance of individual segments of an XR pipeline in terms of end-to-end latency, energy consumption, and AoI. In wireless networks, the mobility of XR devices causes unique transmission and HO delays. Moreover, information from heterogeneous sensors and devices sent to the XR device produces an additional load on the buffer, introducing delays in the overall latency and an increase in energy consumption. In addition, external sensors and devices generate information at their own frequencies, which may cause improper arrival of information packets in the XR pipeline. Our proposed model becomes comprehensive by taking all these details into account, which have not been considered in existing work. Finally, we validated the proposed analytical model against ground truth and compared it with state-of-the-art models. The evaluation shows that our proposed performance analysis modeling framework for XR applications effectively captures and incorporates the important determining factors affecting the end-to-end latency, energy consumption, and AoI, and thus performs better than the compared models with high accuracy.



\end{document}